# ILP-based Alleviation of Dense Meander Segments with Prioritized Shifting and Progressive Fixing in PCB Routing

Tsun-Ming Tseng, Bing Li, Tsung-Yi Ho and Ulf Schlichtmann

*Abstract*—Length-matching is an important technique to balance delays of bus signals in high-performance PCB routing. Existing routers, however, may generate very dense meander segments. Signals propagating along these meander segments exhibit a speedup effect due to crosstalk between the segments of the same wire, thus leading to mismatch of arrival times even under the same physical wire length. In this paper, we present a post-processing method to enlarge the width and the distance of meander segments and hence distribute them more evenly on the board so that crosstalk can be reduced. In the proposed framework, we model the sharing of available routing areas after removing dense meander segments from the initial routing, as well as the generation of relaxed meander segments and their groups for wire length compensation. This model is transformed into an ILP problem and solved for a balanced distribution of wire patterns. In addition, we adjust the locations of long wire segments according to wire priorities to swap free spaces toward critical wires that need much length compensation. To reduce the problem space of the ILP model, we also introduce a progressive fixing technique so that wire patterns are grown gradually from the edge of the routing toward the center area. Experimental results show that the proposed method can expand meander segments significantly even under very tight area constraints, so that the speedup effect can be alleviated effectively in high-performance PCB designs.

## I. INTRODUCTION

DESIGN automation methods for printed circuit boards (PCBs) have advanced steadily in recent years to deal with mounting clock frequencies and increasing complexity of modern PCBs. In a high-performance PCB, interconnects are usually distributed on multiple layers. A wire connecting pins of components may travel across multiple layers to circumvent obstacles and routing congestion. To determine the routing of wires, layer assignment, escape routing, and global routing are usually applied under various constraints. The layer assignment step determines the layers for each wire, while observing given layer constraints and guaranteeing routability with minimum layers, as addressed in [2]–[4]. In the escape routing step, wires are routed from the pins to the boundary of the component [5]–[7]. Moreover, global routing determines the routing of wires between components, while avoiding congestion and maintaining planar routing of buses [8], [9]. These steps, however, may also be combined together for a general solution with better overall routing quality within a short runtime [10].

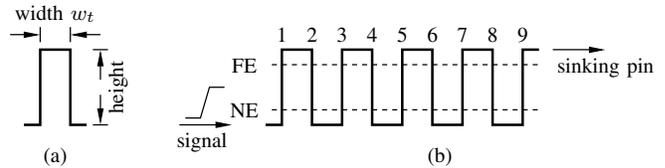

Fig. 1. (a) A meander segment with the width $w_t$. (b) A wire with concatenated meander segments. FE stands for far end and NE for near end.

In modern PCB routers, wires belonging to a bus are usually grouped together, and delay matching between the bus signals has been considered in [11]–[15]. In [11] the delay matching problem is solved by using a Lagrangian model to allocate resources for wire snaking. In [12] this problem is solved with the help of slant symmetric grids. The method in [13] transforms this length matching task to an area assignment problem and proposes a gridless framework using bounded-sliceline grids. The method in [14] routes a given design considering matching wire lengths and wire shapes, while still keeping high efficiency in using routing resources. Specifically, length matching of differential pairs is considered in [15]. Furthermore, the length matching routing problem is also discussed for analog and mixed-signal circuits using detouring shapes in [16], [17], and a longer path algorithm is used in [18] to increase wire delays to meet given specifications.

In the routing methods above, wires that do not have sufficient lengths are extended by creating snaking patterns along them, assuming that signals across wires with the same length have the same delay. These patterns have a high routing density and can be modeled relatively easily, and therefore have gained wide acceptance. Such a snaking pattern is illustrated in Fig. 1a, and is henceforth called a *meander segment*. A wire with concatenated meander segments is shown in Fig. 1b, where NE and FE are abbreviations of near end and far end, respectively.

Accompanying the adoption of the meander segments as a

A preliminary version of this paper was published as [1] in Proceeding of IEEE/ACM International Conference on Computer-Aided Design (ICCAD), 2013.

Tsun-Ming Tseng, Bing Li, and Ulf Schlichtmann are with the Institute for Electronic Design Automation, Technische Universität München (TUM), Munich 80333, Germany (e-mail: tsun-ming.tseng@tum.de; b.li@tum.de; ulf.schlichtmann@tum.de).

Tsung-Yi Ho is with National Chiao Tung University, No. 1001, University Rd., Hsinchu, Taiwan (e-mail:tyho@cs.nctu.edu.tw). The work of T.-Y. Ho was supported in part by the Taiwan Ministry of Science and Technology under grant no. MOST 102-2221-E-009 -194 -MY3, 103-2220-E-009 -029, and 103-2923-E-009 -006 -MY3.

delay compensation method, much research effort has been put into the study of their delay characteristics [19]–[22]. In [19] it is shown that crosstalk noise between meander segments has an accumulation effect in high-speed designs and a *speedup* effect on the wires may appear in such patterns of high density. In [20] a moment technique is proposed to approximate the delays of wires with meander segments, and in [21] a method with finite-difference timing domain is used for qualitative prediction of such delays. Furthermore, in [22] a linear model is formulated to control the wire delay by adjusting the number of meander segments on a fixed-shape wire.

According to [19]–[22], when a signal travels along meander segments, the crosstalk between the segments of the same wire accumulates gradually. Therefore, the signal can reach the sinking pin earlier than estimated using the wire length, thus causing delay mismatch between bus signals. Consider the pattern in Fig. 1b with nine wire segments. At time zero, the main signal switches at the near end of segment 1 and propagates from bottom to top. This signal stimulates crosstalk signals at the near ends of the other wire segments. Assume that the total propagation time of the signal from the near end of one wire segment to the far end of the next wire segment, or from the far end to the next near end, is $t_d$. At time $t_d$, the main signal reaches the far end of wire segment 2 and stimulates a new crosstalk voltage at the far end of the wire segment 3. This new voltage superposes on the crosstalk signal triggered by the main signal on wire segment 2 at time zero, which reaches the far end of wire segment 3 also at $t_d$. This superposition process continues when the main signal propagates across each wire segment, and finally the crosstalk voltage may surpass the threshold of logic switching before the main signal, leading to a speedup effect on the wire.

Despite the crosstalk noise between meander segments, they are still widely used in floorplan-like or area-assignment-like routing methods such as [11]–[14], because they can be applied relatively easily to match wire lengths and can be adjusted flexibly. Other delay matching patterns, for instance, the concentric delay line [21] or the flat spiral delay line [23], impose more computational complexity and are still not widely applied, especially in existing commercial tools.

To alleviate the delay mismatching problem discussed above, we present in this paper a mathematical model to generate meander patterns in a routing. We also propose an iterative algorithm to find the largest possible width $w_t$ for the newly created meander segments, which are used to compensate wire lengths after we remove the dense meander segments from the original routing. The resulting routing has a similar shape and the same wire lengths, but with meander segments of enlarged width distributed evenly in the routing. After applying the proposed method, wire delays can be estimated by wire lengths more accurately due to the alleviation of the speedup effect. In addition, the proposed method adjusts routing results from other routers, which have already determined the basic routing patterns and wire lengths, so that it can be integrated into an existing PCB design flow seamlessly.

The rest of the paper is organized as follows. In Section II we give the formulation of the problem to be solved in this paper. In Section III we describe the proposed ILP-based model of routing patterns for the alleviation of dense meander segments in details. Thereafter, we explain the iterative prioritized shifting of long straight wire segments and the progressive fixing of patterns in Section IV. We show experimental results in Section V and conclude the paper in Section VI.

## II. Overview

In this paper, we try to alleviate the speedup effect by increasing the width of meander segments $w_t$ as illustrated in Fig. 1a, i.e., the distance between the wire segments forming such a pattern. First, we delete existing dense meander segments from the original routing to create a starting floorplan, as explained in Section III-A. Thereafter, we model the minimum distance between the segments of a wire with the variable $w_t$. Then, a model for routing patterns containing multiple wires and sharing the same free space is established. The modeling of wire patterns is explained in Section III-B. This model is transformed into an ILP problem and processed by a solver to generate the routing for a given $w_t$. The largest value of $w_t$ in our formulation is then found by binary search with several iterations, as explained in Section III-C. During these iterations, long wire segments are shifted to move free spaces toward wires that need more routing space. In addition, wire patterns are gradually fixed from the edge of the routing area to reduce the problem space. These techniques are applied in the iterations described in Section III-C, and their detailed explanation can be found in Section IV.

In the proposed method, we model that all meander segments share the same minimum distance $w_t$, so that the speedup effect can be addressed at the routing level. Namely, we can reduce this effect by simply enlarging the shared width $w_t$ in the routing. If, otherwise, different widths $w_t$s of meander segments are allowed, or different routing patterns such as the concentric delay line [21] or the flat spiral delay line [23] are used to extend wire length, a close look at the result of signal simulation might be inevitable for an accurate delay evaluation. This simulation is timing-consuming and the resulting routing requires an accurate control in manufacturing to match the parameters used in simulation.

The problem formulation of this paper is as follows:
*input*: a PCB routing or a floorplan with given wire lengths.
*output*: a routing with $w_t$ as the width of meander segments.
*objective*: to enlarge $w_t$ as much as possible.
*constraints*: all wires should keep their original lengths or meet the given length specifications, and new meander patterns should not exceed the routing boundaries. These boundaries are defined by flattening the outermost wires in the bus and the other neighboring wires if they have routing conflicts with the flattened wires.

## III. Alleviation of Meander Segments using an ILP Model

In this section we describe an ILP model for patterns in a routing area and space sharing by multiple wire groups. In addition, an iterative algorithm is introduced to find the largest possible width $w_t$ for the newly created meander segments, while complying with given area constraints and guaranteeing that the distance between any two wires is larger than $d_m$ required for manufacturing.

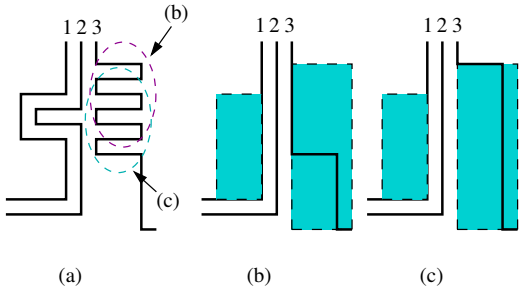

Fig. 2. Removal of dense meander segments. (a) Original routing with dense meander segments. (b)/(c) Dense meander segments are removed under different searching directions.

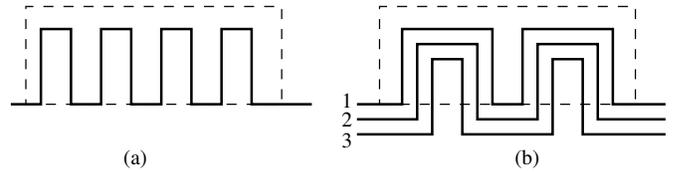

Fig. 3. Growth of meander segments in a free space. (a) Single wire. (b) Wire groups with three wires.

### A. Removal of dense meander segments

The first step of the proposed method is to remove dense meander segments from the initial routing to create a routing floorplan. New meander segments with a minimum width $w_t$ will be grown in the free spaces available thereafter. Fig. 2 shows an example of deleting dense meander segments with widths no larger than three units. The shaded areas with dashed boundaries are the available free spaces for growing new meander segments. In this operation, the dense meander segments on wire 1 and 2 can be removed simply, but the ones on wire 3 can be viewed from different sides of this wire, so that both floorplans in Fig. 2b and 2c can be generated.

Although the free space in Fig. 2b is split into more parts, we cannot assert that Fig. 2c is *globally* better than Fig. 2b, because the relations of wires and all free spaces along the bus determine the best configuration together. In the proposed method, we trace the wires from one direction to delete meander segments. For example, if we trace the wires from top to bottom in Fig. 2a, we can identify the free space as in Fig. 2b. Comparing these cases, we notice that the case in Fig. 2c can be formed from Fig. 2b by shifting the horizontal wire segment that splits the free space upwards. In the proposed method, we first try to grow new dense meander segments in the available free spaces created by one-directional tracing. If some wires need more space according to the results from the first iterations, the straight wire segments such as in Fig. 2b are shifted in later iterations to make more routing area available to critical wires. This technique will be discussed in Section IV in more detail.

### B. Growth of meander segments and space sharing

In this section we explain the modeling of meander segments in a given free space and the sharing of a free space by multiple wire groups. The generated constraints will be used by the algorithm in Section III-C to find an optimal solution.

*1) Modeling the patterns in a given free space*

In a given free space the growth of meander segments with minimum width $w_t$ on *one* wire can be performed relatively easily by calculating the allowed number of meander segments in this area. Fig. 3a shows such an example. In the routing of a bus, however, more than one wire can exist at a side of a free space. For example, three wires at the bottom of the free space in Fig. 3b can form meander segments in the free space at the same time. With this pattern, wires that are below the other wires still have a chance to use the free space to extend their lengths, but at the expense that the widths of the meander segments of the upper wires should be increased, so that fewer meander segments for the outer wires to compensate the removed wire segments can be created. Comparing Fig. 3a and 3b, we can observe that this is a tradeoff between sharing the free space with a group of wires and maximizing the length compensation of individual wires. The complexity comes from the fact that it is not easy to determine which wires should be pushed into the free space, and how many meander segments should be formed on each wire. In addition, there exists a dependency between these newly formed meander segments. For example, wire 3 can be pushed into the free space only when wire 1 and 2 are pushed upwards.

To solve the problem described above, we formulate a flexible model to handle the number of meander segments in the free space and the dependency between multiple wires. A general analysis of the new meander segments in a free space is shown in Fig. 4, where four upward meander segment groups (msg) are illustrated to show different relations of the wires. In $msg_1$ the lower meander segment has enough height to take a part of the internal space of the upper segment, so that the width of the upper segment must be at least two times of the minimum wire distance $d_m$ larger than the width of the lower meander segment. In $msg_2$ only the left vertical wire segment of the upper meander segment has conflict with the lower meander segment so that the width of the upper meander segment needs only to be increased by one $d_m$. In $msg_3$, no conflict exists so that both meander segments can have the width $w_t$. In the last group the lower wire does not have a meander segment because no wire length compensation is needed. In this case, the width of the upper meander segment is equal to $w_t$.

In addition to the segment groups in Fig. 4, intermediate upward segments may be grown, for example, the one shown with a dashed line. In modeling all the relations in the meander segment groups, the intuitive model in Fig. 4 requires variables to be assigned for all the heights of the meander segments, and to describe the possible relations of the wires with many constraints. For example, the widths of meander segments on wire 1 depend on the heights of those on wire 2. However, not all the flexibility provided in Fig. 4 is really necessary, because different wire heights may lead to inefficiency in using free spaces, for example, the free space in $msg_1$ and $msg_3$ and above $msg_2$ may be wasted. When the model is processed by a solver, such cases are rarely selected because the objective of optimization includes the maximization of wire length compensation and the usage of free spaces.

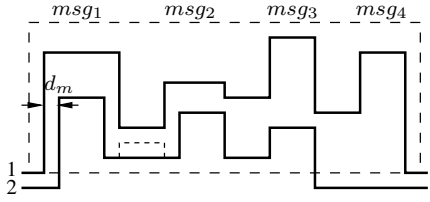

Fig. 4. General model of meander segments of different heights.

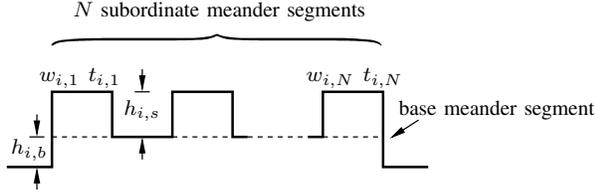

Fig. 5. Simplified meander segment model of a wire described with fewer variables.

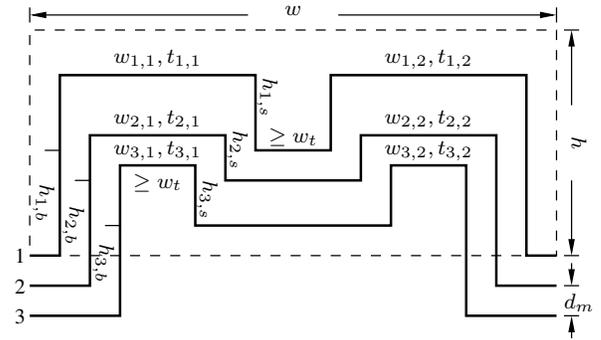

Fig. 6. Simplified model of meander segments for a wire group with pattern dependency.

Based on the general formulation in Fig. 4, we propose a simplified model with fewer variables and constraints. The variables for the $i$th wire are shown in Fig. 5. This model can be considered as two parts, where $N$ *subordinate meander segments* with heights $h_{i,s}$ grow at the top of the *base meander segment,* which has the height $h_{i,b}$ and spans across all the horizontally available space. In this model, the subordinate meander segments increase the length of the $i$th wire, while the base meander segment shifts the free space so that the other wires below it also have access to the free space. To model the heights of the pattern in Fig. 5, we need only two variables $h_{i,b}$ and $h_{i,s}$, instead of the $2N$ variables if the general formulation in Fig. 4 is used. In Fig. 5, the width of the $j$th subordinate meander segment of the $i$th wire is represented using $w_{i,j}, j = 1, \ldots N$. In addition, we assign a 0-1 variable $t_{i,j}$ for each subordinate meander segment in Fig. 5 to model its appearance in the final routing. These variables will be discussed later in this section.

This simplified model sacrifices the flexibility in selecting heights of the subordinate meander segments. However, this flexibility does not contribute to the maximization of wire length compensation much because the irregular heights in Fig. 4 may actually waste the free space in growing meander segments as discussed before. Though we do not consider all the possible patterns in the simplified model, for example, the intermediate meander segments shown with the dashed line in Fig. 4, directly, these possibilities are explored by the iterations in the proposed method in Section III-C, where new wire patterns are grown in the free spaces left by the model in earlier iterations.

With the simplified model in Fig. 5, we can now model the growth of meander segments from a group of wires into a free space, as shown in Fig. 6, where three wires and two subordinate meander segments on each wire are used as example. Each wire in Fig. 6 has the variables assigned as in Fig. 5. For the $i$th and $(i+1)$st wires, the distance between the horizontal wire segments should always be larger than the minimum wire distance $d_m$ to guarantee manufacturability.

These constraints can be described as

$$(h_{i,b} + h_{i,s} + d_m) - (h_{i+1,b} + h_{i+1,s}) \geq d_m \quad (1)$$
$$(h_{i,b} + d_m) - h_{i+1,b} \geq d_m \quad (2)$$

where $i = 1, \ldots M - 1$. $M$ is the number of wires and equal to 3 in Fig. 6. Here we have assumed that the wire distance in the original wire group is $d_m$ to simplify the expressions in (1) and (2). The constraint (1) describes the relations of horizontal segments at the top of the subordinate meander segments, and (2) the relations at the bottom of the subordinate meander segments. From (1)–(2), we can observe that the subordinate meander segments on the first wire have the largest height. In order to fit the grown patterns into the free space, this height should be no larger than the height of the given free space, that is,

$$h_{1,b} + h_{1,s} \leq h \quad (3)$$

where $h$ is the height of the free space as shown in Fig. 6.

As mentioned earlier, the 0-1 variable $t_{i,j}$ defines whether a subordinate meander segment can exist in the final routing. Examining the patterns in Fig. 6, we can find that a subordinate meander segment can only grow upwards when its upper neighboring wire has such a pattern. Assume that $t_{i,j} = 1$ when the $j$th subordinate meander segment on the $i$th wire exists. The vertical dependency constraint can be modeled as

$$t_{i,j} \geq t_{i+1,j}, i = 1, \ldots M - 1. \quad (4)$$

According to this constraint, if a subordinate meander segment does not exist, the ones below it cannot be created to extend wire lengths. That is to say, all the subordinate meander segments below it are blocked due to the nature of the chained constraint (4). In Fig. 6 the widths of the meander segments should be no smaller than the given minimum distance $w_t$. Therefore, we can model the constraints for the width $w_{i,j}$ of the $j$th subordinate meander segment on the $i$th wire as

$$w_{i,j} \geq t_{i,j} w_t + \sum_{k=i+1}^{M} 2 t_{k,j} d_m \quad (5)$$

where $d_m$ is the minimum space between wires and $\sum_{k=i+1}^{M} 2 t_{k,j} d_m$ is the sum of increased widths due to the meander segments which are surrounded below the $i$th wire.

From Fig. 6 we can observe that the width of the complete pattern is bounded by the width of the first wire. This width must be no larger than the width of the given free space, so that we have

$$\sum_{k=1}^{N} w_{1,k} + \sum_{k=2}^{N} t_{1,k} w_t \leq w \quad (6)$$

where $w$ is the width of the given free space shown in Fig. 6, and $\sum_{k=1}^{N} w_{1,k}$ is the sum of the widths of the subordinate meander segments on the first wire as constrained in (5). $\sum_{k=2}^{N} t_{1,k} w_t$ is the sum of the distances between the subordinate meander segments on the first wire. Each of these distances should be no smaller than the minimum width $w_t$ of meander segments, as shown by $\geq w_t$ on the first wire in Fig. 6. Note here the constraint (6) requires that the meander segments should be enabled from index 1 forth. This prerequisite will be discussed soon in this section.

With the 0-1 variables $t_{i,j}$, we provide more freedom to the solver to select how many meander segments should be created in the free space and how many wires should be pushed together. Consider the cases illustrated in Fig. 3. If wires tend to be pushed into the free space as a group as in Fig. 3b, the number of meander segments becomes smaller than in the case that only a few wires are pushed into the free space, as in Fig. 3a. Here exists a tradeoff between sharing the free space among a group of wires and maximizing the length compensation of individual wires. When we model the possible patterns in a free space, we do not know how many meander segments should be created to achieve an optimal solution. But this number has an upper bound, which can be computed as the number of meander segments when only one wire is pushed, similar to the case in Fig. 3a. With this analysis, we compute the maximum number of possible meander segments in a free space as

$$N = \lfloor (w - w_t)/2w_t \rfloor + 1 \tag{7}$$

where $w$ is the width of the free space and the symbol $\lfloor \ \rfloor$ represents the greatest integer no larger than the parameter. Note that $N$ is the maximum number of possible meander segments.

If a group of wires form meander segments together in the free space, that is, they are pushed together into the free space similar to Fig. 3b, the number of meander segments drops significantly. Consider a wire group with $M$ wires. The maximum number of meander segments $N_1$ for the first wire can be calculated using (7), so that $N = N_1$, but the $M$th wire cannot have as many meander segments, because creating a meander segment on the $M$th wire requires that all the wires above should be pushed upwards. Therefore, fewer meander segments can be created in the free space. For example, in Fig. 3b, wire 3 can have at most two meander segments. With this observation, we can reduce the number of 0-1 variables $t_{i,j}$ for different wires. If a meander segment on the $i$th wire in a wire group should be created, the minimum width of the corresponding meander segments on the first wire can be calculated using (5). In addition, as defined by (4), if the $i$th wire is pushed upwards, $t_{i,j}$ should be set to 1 and thus $t_{k,j}, k = 1, \ldots i-1$ should also be 1. Therefore, the width of the uppermost meander segment under this condition should meet

$$w_1^i = w_t + \sum_{k=2}^{i} 2d_m \leq w_{1,j}, \ i = 2, \ldots M. \tag{8}$$

Similar to (7), the maximum number of meander segments $N_i$ for the $i$th wire can be computed as

$$N_i = \lfloor (w - w_1^i)/(w_1^i + w_t) \rfloor + 1. \tag{9}$$

Comparing (7) and (9), we can find that $N_i$ may be much smaller than $N$ for the $i$th wire. To reduce the number of variables, we set the last $N - N_i$ 0-1 variables for the $i$th wire to zero, as $t_{i,j} = 0, j = N_i + 1, \ldots N$, because these meander segments will never be feasible.

In the discussion above, we expect the solver to grow meander segments at the first $N_i$ positions for the $i$th wire. In order to align the meander segments on different wires so that they can be formed into a group as in Fig. 6, we add additional constraints for all the wires as

$$t_{i,j} \geq t_{i,j+1}, \ i = 1, \ldots M, \ j = 1, \ldots N-1. \tag{10}$$

These constraints force the solver to create the meander segments from the beginnings of the wires. If a meander segment is not enabled, all the following ones on the same wire cannot be enabled either. In fact, adding these constraints does not affect the compensated wire lengths, because the new constraints simply rearrange the order of the freely selected meander segments. These constraints are also required by (6) for calculating the total width of a wire pattern.

In forming meander segments in free spaces, we increase the lengths of wires to compensate the meander segments removed from the original routing as described in Section III-A. For the $i$th wire in Fig. 6, the compensated length can be expressed as

$$l_i = l_{i,b} + \sum_{k=1}^{N_i} 2t_{i,k} h_{i,s} \tag{11}$$

where $\sum_{k=1}^{N_i} 2t_{i,k} h_{i,s}$ is the sum of the heights of all the subordinate meander segments which appear in the final routing with $t_{i,k} = 1$. $l_{i,b}$ is newly defined here to represent the wire length increased by the height of the base meander segment as discussed earlier using Fig.5. If there exist any subordinate meander segments on the $i$th wire, the base meander segment is always included in the model to shift space to the wires below, so that $l_{i,b} = 2h_{i,b}$ as illustrated in Fig. 6; otherwise $l_{i,b}$ is equal to zero. This description is equivalent to

$$\text{if } \sum_{k=1}^{N} t_{i,k} \geq 1, \text{ then } l_{i,b} = 2h_{i,b}; \text{ else } l_{i,b} = 0. \tag{12}$$

Considering the constraints (10) we can find that the condition $\sum_{k=1}^{N} t_{i,k} \geq 1$ in (12) is equivalent to $t_{i,1} = 1$, because any $t_{i,j} = 1, j = 2, \ldots N$ requires that $t_{i,1} = 1$. Therefore, we can transform the constraint in (12) to

$$\text{if } t_{i,1} = 1, \text{ then } l_{i,b} = 2h_{i,b}; \text{ else } l_{i,b} = 0 \tag{13}$$
$$\iff l_{i,b} = 2t_{i,1} h_{i,b}. \tag{14}$$

Using (11) and (14), we can express the increased length of a wire using the sum of products of a 0-1 variable with either the base height $h_{i,b}$ or subordinate height $h_{i,s}$. This formulation is in a quadratic form because it contains the sum of $t_{i,k} h_{i,s}$ and $t_{i,1} h_{i,b}$ where all the values of $t_{i,k}$, $h_{i,s}$, $t_{i,1}$ and $h_{i,b}$ should be determined by the solver. We will explain the transformation of this formulation into an ILP problem later.

The model in Fig. 6 describes the pattern in one free space. However, it is very common that a group of wires has free spaces on both sides of it. In this case, some wires may form meander segments in the upper space and others may use the lower space, as illustrated in Fig. 7. As we have discussed before, if more wires are pushed into a free space, fewer meander segments can be formed. Therefore, we should allow the solver to select the direction of the meander segments on each wire. To meet this requirement, we assign a new 0-1 variable $t_i$ for the $i$th wire. If the meander segments on this wire are upward, $t_i$ is set to 1; otherwise, $t_i$ is set to 0. Because a wire can be pushed upwards only when the wires above it are pushed upwards, we can establish the relation between the new variables as

$$t_i \geq t_{i+1}, i = 1, \ldots, M-1 \quad (15)$$

which also implies the downward constraints.

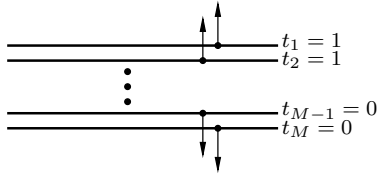

Fig. 7. Selecting directions of meander segments.

Now consider the patterns in Fig. 6. If the solver determines the $i$th wire should form upward patterns, this wire must be allowed to be pushed upwards, which means

$$t_{i,j} \leq t_i, j = 1, \ldots, N. \quad (16)$$

If the $i$th wire is pushed downwards, the corresponding constraints are written as

$$t_{i,j} \leq 1 - t_i, j = 1, \ldots, N. \quad (17)$$

Here $t_{i,j}$ indicates whether a pattern can be formed on a wire, either upwards or downwards. This direction is actually controlled by $t_i$ in (17). For example, if the solver selects the last two wires in Fig. 7 to be pushed into a downward space, the variables $t_i$ ($t_{M-1}$ and $t_M$ in Fig. 7) are set to 0, so that $1 - t_i = 1$. Consequently, $t_{i,j}$ has the chance to be set to 1 to grow wire patterns.

In finding an optimal solution, the solver can determine which wires should be pushed upwards or downwards. That is, the wires are partitioned to two groups automatically by setting the values $t_i, i = 1, \ldots, M$, so that a proper number of meander segments $N_i$ as defined in (9) for each wire can be chosen to establish a balanced length compensation on all wires. Note here we assign only one variable $t_i$ for the $i$th wire, so that all the meander segments on a wire should be pushed upwards or downwards at the same time. We do not allow individual selection of the direction of each meander segment, so that the number of variables can be reduced. If we find that some wires cannot be compensated very well in the iterations, we shift all the wire groups downwards or upwards to swap or merge the two free spaces as to be discussed in Section IV.

*2) Modeling the sharing free space by multiple wire groups*

In the last section, we have explained how to model the meander patterns in a free space. After the original dense meander segments are removed, there may be some free spaces surrounded by different wire groups. All these wire groups may grow meander segments into the same space, leading to a resource sharing problem. Fig. 8 shows an example of space sharing. In this example, the free space $S$ represented by the rectangular area is shared by four wire groups. For a wire group $wg_i$, a free space $S_i$ is allocated from $S$ to grow meander segments. But the dimensions of $S_i$ should be determined considering the relations of other wires and required length compensation globally. In this paper, we use a model also based on the 0-1 selection variables to describe the possible combinations of allocated free spaces for wire groups, and an ILP solver is used to determine an optimal solution.

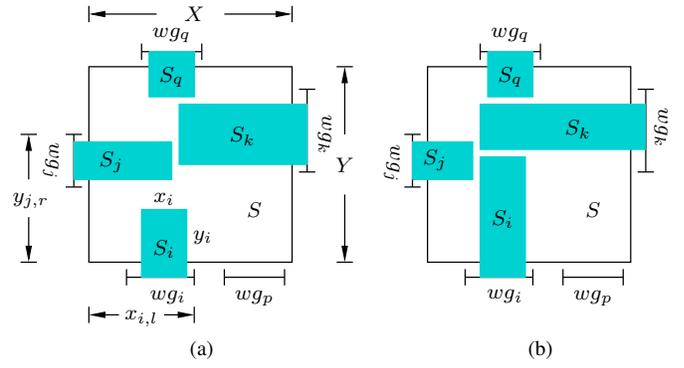

Fig. 8. Model of sharing free space by multiple wire groups. (a) $c_{i,j} = 0$ so that $wg_i$ is blocked by $wg_j$. (b) $c_{i,j} = 1$ and $c_{i,k} = 0$.

We use the wire group $wg_i$ in Fig. 8 as an example to explain the model for space sharing. From $wg_i$ upwards, the free space $S_i$ may have conflicts with the free spaces growing in the horizontal direction. For example, $S_j$ blocks $S_i$ in Fig. 8a. In Fig. 8b, $S_i$ blocks $S_j$ but is blocked by $S_k$. To model these conflict conditions, we assign a 0-1 variable $c_{i,j}$ for the wire group pair $(wg_i, wg_j)$. If $c_{i,j} = 1$, $wg_i$ can pass $wg_j$; otherwise, $wg_i$ is blocked by $wg_j$. For example, $c_{i,j} = 0$ in Fig. 8a, and $c_{i,j} = 1, c_{i,k} = 0$ in Fig. 8b.

For the wire group $wg_i$, we represent its wire length using $w_i^c$, which is a constant and set after the original dense meander segments are removed. For the free space $S_i$, we describe its horizontal dimension as $x_i$ and its vertical dimension as $y_i$, as shown in Fig. 8a. Note that $x_i$ and $y_i$ both are variables and should be set by the solver. Because the wires can only be pushed in their perpendicular direction, the dimension $x_i$ of the free space $S_i$ should meet

$$x_i \leq w_i^c. \quad (18)$$

Similarly, a constraint for a wire group $wg_j$ at the left side of the free space $S$ can be written as

$$y_j \leq w_j^c. \quad (19)$$

Because the dimensions of the free spaces for the wire groups, for example, $x_i$ and $x_j$, are all variables, the solver has the freedom to balance the area usage between different wire groups. For example, in Fig. 8a the vertical dimension of the free space for $wg_j$ can be set relatively small, so that the free space for $wg_i$ can be enlarged for more length compensation.

The next set of constraints are from the overlapping wire groups at either the horizontal direction or the vertical direction. For example, in Fig. 8 $wg_i$ and $wg_q$ overlap horizontally, and $wg_j$ and $wg_k$ overlap vertically. If free spaces are established from both wire groups in an overlapping pair, the corresponding dimensions should be constrained so that the total dimension of the free space $S$ is not exceeded. For example, the overlapping group pairs $(wg_i, wg_q)$ and $(wg_j, wg_k)$ have the dimensional constraints

$$x_j + d_m + x_k \leq X \quad (20)$$
$$y_i + d_m + y_q \leq Y. \quad (21)$$

Now we establish the constraints from the blocking combinations of wire groups. Using wire group $wg_i$ as an example,

we search upwards from it to establish the blocking constraints with other wire groups at the left or the right side of $S$. At first we can find that if $wg_i$ passes a wire group $wg_k$, it should pass the wire group $wg_j$ which is below $wg_k$. If we order the wire groups at the left and right sides of $S$ according to their positions from bottom to top, we can deduce the dependency between the 0-1 variables assigned before as

$$c_{i,j} \leq c_{i,j-1} \tag{22}$$

where the $(j-1)$th wire group is below the $j$th wire group.

In Fig. 8a, if $wg_i$ is blocked by the wire group $wg_j$ from the left side, that is, $wg_j$ spans above $wg_i$, the vertical dimensions of their free spaces $S_i$ and $S_j$ should meet

$$y_i + d_m + y_j \leq y_{j,r} \tag{23}$$

where $y_i$ and $y_j$ are the dimensions of $S_i$ and $S_j$ in the vertical direction, respectively. $d_m$ is the minimum space between wires. $y_{j,r}$ is the distance from the bottom of $S$ to the left end of $wg_j$. This constraint is only valid when the 0-1 variable $c_{i,j}$ is 0, and we can incorporate this additional constraint into (23) as

$$y_i + d_m + y_j + (1 - c_{i,j})\Gamma \leq y_{j,r} + \Gamma \tag{24}$$

where $\Gamma$ is a predefined very large constant. If $c_{i,j}$ is equal to 0, the constraint is the same as (23). If $c_{i,j}$ is equal to 1, the constraint becomes

$$y_i + d_m + y_j \leq y_{j,r} + \Gamma \tag{25}$$

which always holds for a very large $\Gamma$. With this technique, we can now express the other constraints in the space sharing. The constraint (24) is valid when $wg_i$ is blocked by $wg_j$. If $wg_i$ can pass $wg_j$, as shown in Fig. 8b, the width relation should be established as

$$x_j + d_m + x_i + c_{i,j}\Gamma \leq x_{i,l} + \Gamma \tag{26}$$

where $x_{i,l}$ is the distance from the left side of $S$ to the right end of $wg_i$. This constraint is only relevant when $c_{i,j} = 1$; otherwise it always holds and has no effect on the model. If we search from $wg_i$ upwards further, $wg_i$ may block all the wire groups from left and right sides. In this case the sum of the corresponding vertical dimensions of $wg_i$ and the wire groups at the top which $wg_i$ overlaps should not exceed the height $Y$ of $S$, as already defined in in (21).

The constraints (20)–(26) are created for the wire group $wg_i$ at the bottom side. For the wire groups at the top of $S$, for example, $wg_q$, we need to establish their conflict constraints similar to (20)–(26) downwards. Note here we do not establish the relations of the 0-1 variables from a wire group at the top and from a wire group at the bottom. If the solver allows both wire groups to block the same wire groups from left or right sides, some area overlap between the upper group and the lower group seems to appear. However, the constraints (20)–(21) guarantee that there is no such overlap in the final allocation of free spaces for wire groups, and the solver will select an optimal relation between their dimensions to maximize the length compensation.

The horizontal and vertical dimensions in Fig. 8 describe the available free spaces allocated to wire groups. For example, $x_i$ and $y_i$ are the width $w$ and height $h$ in (6) and (3), respectively. With this connection, the relation between the allocation of free spaces for wire groups and the number of possible meander segments as well as the reintroduced wire lengths in Fig. 6 is established and the solver has the freedom to select an optimal solution from the configurations for all the wire groups around the free space.

In real routing, the free space after removing the original meander segments may be irregular. When identifying these spaces, we try to determine the largest rectangular area for each wire group. For example, in Fig. 8 the wire groups $wg_i$ and $wg_p$ may have different available heights $Y$ because they may not be aligned in the original routing. This identification process, however, still leaves some free spaces unused. In the next section, we will introduce an iterative algorithm to improve the efficiency of space usage.

### C. Solving the model and iterative length compensation

Till now we have discussed the constraints in an available free space and the sharing of free spaces between different wire groups. In this section, we will formulate the ILP problem and explain an iterative algorithm to compensate the length of meander segments.

In Section III-B1, the proposed model works with a given width of meander segments. To find the largest possible width of these meander segments, we apply a binary search. In each iteration, a minimum width of meander segments $w_t$ is given, and an optimization problem is formulated as what follows. From Fig. 6, we can see that the compensated wire lengths come from the heights of newly created meander segments. For a free space, the sum of compensated lengths of a wire is defined in (11). Assume that we have in total $M_t$ wires in the design and the $j$th wire belongs to $n_j$ wire groups when creating the patterns in free spaces. We write the indexes of this wire in the $n_j$ groups as $i_1, i_2, \ldots i_{n_j}$, respectively. Therefore, the total compensated length for the $j$th wire can be computed using (11), as

$$L_j = \sum_{k=1}^{n_j} l_{i_k}. \tag{27}$$

After removing the dense meander segments from the original routing, we know how much wire length we should compensate to maintain the same length for each wire as in the original routing. Such a length for the $j$th wire is a constant and represented by $L_j^c$. To guarantee the same wire lengths, we need the following constraints,

$$L_j = L_j^c, j = 1, \ldots M_t. \tag{28}$$

Therefore, the goal of the optimization is to find a solution for all the variables involved in the constraints established in Section III-B1 and III-B2 so that (28) holds.

In the optimization problem above, the compensated wire lengths contains quadratic terms, for example, $t_{i,k}h_{i,s}$ in (11). In this term $t_{i,h}$ is a 0-1 variable so that we can convert all these quadratic terms to linear form. By changing the index $i$ to $i_k$ and setting $z_{i_k,v} = t_{i_k,v}h_{i_k,s}, z_{i_k,v} \geq 0$, we can transform (11) into a linear form

$$l_{i_k} = l_{i_k,b} + \sum_{v=1}^{N} 2z_{i_k,v}. \tag{29}$$

According to [24], the new constraint $z_{i_k,v} = t_{i_k,v}h_{i_k,s}$ can be split into $z_{i_k,v} \geq t_{i_k,v}h_{i_k,s}$ and $z_{i_k,v} \leq t_{i_k,v}h_{i_k,s}$. These

**Algorithm 1:** Progressive alleviation of meander segments using an ILP model with iterations

```
L1   Input:
L2      R_i: initial routing;
L3      w_t: the given minimum width of meander segments;
L4   Output:
L5      R_o: improved routing;
L6   Variables:
L7      R: the current routing needing length compensation;
L8      M: an ILP model from a given routing R;
L9      S: a solution of the ILP model M;
L10     W: a set of wires with lengths to be compensated;
L11     w_t: the minimum width of meander segments;
L12     \overline{w_t}: the upper bound of w_t;
L13     \underline{w_t}: the lower bound of w_t.
L14  R_o ← R_i;
L15  w_t = \overline{w_t};
L16  repeat
L17     R ← remove_meander_segments (R_i, w_t);
L18     initialize_priorities (R);
L19     for i=1 to n_s do
L20        shift_segments (R);
L21        for j=1 to n_r do
L22           M ← create_ILP_model (R, w_t);
L23           S ← ILP_solve (M);
L24           R ← grow_patterns (R, S);
L25           W ← critical_wires (S);
L26           if W is empty then
L27              R_o ← R;
L28              \underline{w_t} = w_t;
L29              w_t = (\underline{w_t} + \overline{w_t})/2;
L30              go to L38;
L31           end
L32        end
L33        update_priorities (R);
L34        fix_remove_patterns (R, w_t);
L35     end
L36     \overline{w_t} = w_t;
L37     w_t = (\underline{w_t} + \overline{w_t})/2;
L38  until \overline{w_t} − \underline{w_t} < step;
L39  return R_o;
```

split constraints can be transformed into linear forms as

$$z_{i_k,v} \geq t_{i_k,v} h_{i_k,s} \iff z_{i_k,v} \geq h_{i_k,s} - (1 - t_{i_k,v})\Gamma \quad (30)$$

$$z_{i_k,v} \leq t_{i_k,v} h_{i_k,s} \iff z_{i_k,v} \leq h_{i_k,s} \text{ and } z_{i_k,v} \leq t_{i_k,v}\Gamma \quad (31)$$

where $\Gamma$ is a predefined very large constant. These transformations can be verified by checking the equivalence when $t_{i_k,v}$ is set to 0 or 1 individually. Similarly, we can also transform the quadratic terms in (14) and accordingly $l_{i_k,b}$ in (29) to linear forms so that the compensated wire length $L_j$ in (27) is converted into a linear form.

The model explained in Section III-B1 and III-B2, though flexible, still cannot cover all the possibilities to grow new meander segments because it only models patterns in Fig. 6 and generates them on only one side of a wire segment and thus may still leave free spaces in the routing. To improve the efficiency of area usage, we run the modeling and solving process by several iterations, each of which uses the result of the last one as input. Therefore, the unoccupied free spaces are used again in the following iterations to create an ILP model for new patterns, leading to a much higher routing density. With this concept, we try to maximize the compensated wire lengths in each iteration. Instead of using the constraints in (28), with which the solver may simply report an infeasible solution in the first iteration and produce no result for further iterations, we use the following constraints

$$L_j \leq L_j^c, j = 1, \ldots M_t. \quad (32)$$

Therefore the new optimization problem can be described as

$$\text{maximize: } \sum_{j=1}^{M_t} L_j \quad (33)$$

$$\text{subject to: all } linear \text{ constraints in} \quad (34)$$

Section III-B1 to III-C except (28).

This formulation is an ILP problem and can be solved directly.

The pseudo code of compensating wire lengths using the ILP model is shown in Algorithm 1. The input of this algorithm is the original routing, denoted by $R_i$. The output of this algorithm is the improved routing $R_o$ containing meander segments with an enlarged width $w_t$. The wires in the improved routing have the same lengths as in the original routing with dense meander segments. The loop L21–L32 creates the ILP model and calls the solver $n_r$ times. Inside each of these $n_r$ iterations, new meander segments are grown incrementally in the routing $R$ in the free spaces left by the last iteration. These iterative improvement steps are used to overcome the limitation of the ILP model.

In any of the $n_r$ inner iterations L21–L32, if no critical wire exists in the result as checked in L26, meaning that all wire lengths are compensated, a feasible routing has been found for a given width $w_t$. If, however, the wire lengths cannot be fully compensated after these $n_r$ inner iterations, the three functions in L20, L33, and L34 are used to adjust the floorplan according to the result of the wire length compensation. These outer iterations are executed $n_s$ times to check the feasibility of the current width $w_t$. In addition, all the iterations are included in a binary search procedure in L16–L38 to find a $w_t$ as large as possible for the given routing. In the next section, we will explain the floorplan adjustment functions in L20, L33, and L34 of Algorithm 1 in detail.

## IV. PRIORITY-BASED SEGMENT SHIFTING AND PROGRESSIVE PATTERN FIXING

The proposed ILP model can grow meander segments only to one side of a long straight wire segment and Algorithm 1 starts from only one of the possible floorplan variants such as in Fig. 2b or 2c. These simplifications may lead to a result that uses the routing area inefficiently. To overcome these limitations while still keeping a reasonable runtime, we apply prioritized shifting and progressive fixing techniques to adjust the floorplan iteratively.

### A. Shifting long straight wire segments according to priorities

To mitigate the problem of split free spaces, we can change the positions or shift the straight wire segments to merge the free spaces on both sides. For example, if the rightmost vertical wire segment in Fig. 2b or 2c can be pushed leftwards to form the floorplan in Fig. 9a, the two free spaces are merged into one and become available to the third wire segment. Furthermore, this shift operation forms a wire group containing three wires. Therefore, complex patterns similar to Fig. 3b may be grown, so that the free space is also available to the other two wires by group patterns.

While shifting straight segments may assign more routing resource to critical wires, a global optimal shift mechanism cannot be achieved easily for several reasons. First, the option to enumerate all possible floorplans from wire shifting can be ruled out due to their large number. Second, the shift directions of wire segments depend on the amount of wire lengths to be compensated. These lengths are only available after running the ILP solver, which grows new meander segments from the result of the last iteration so that a direct accurate model for shifting is not easy to form.

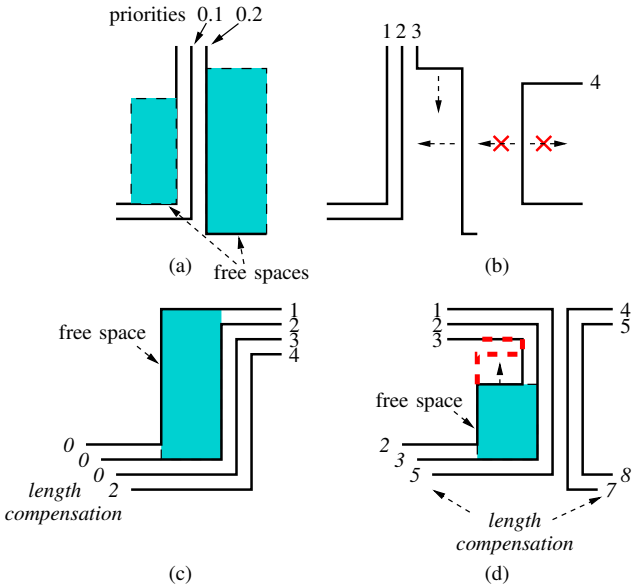

Fig. 9. Segment shift operations. (a) Merging free spaces by moving the rightmost vertical segments in Fig. 2a/b. (b) Shifting straight segments without changing wire length. (c) Priorities determined by the uncompensated lengths of multiple wires accumulatively. (d) Shifting segments and deleting dense meander segments with wire length change.

To deal with the challenge in shifting straight wire segments, we employ an iteratively adaptive algorithm, following the nature of the iterative optimization in Algorithm 1. In this method, we assign priorities to wires according to lengths uncompensated by the optimization in the last iteration. Because the wires with large uncompensated lengths need more space and flexibility in the following iterations, their priorities are raised accordingly. Thereafter, wire segments are shifted according to these priorities in two criterion. First, wire segments are shifted to the neighboring segments with smaller priorities. Second, the shift operation should not change the lengths of the moved wires. Otherwise, we should update the priorities of these wires accordingly, but this may lead to a manual routing and thus non-optimal result. In Fig. 9b we show possible segment shift operations. In this example, both the horizontal and vertical segments of wire 3 can be shifted in the directions of the dashed arrows. However, the segments on wire 4 cannot be moved leftwards or rightwards, because these operations change the length of wire 4.

The quality of the results produced by shift operations depends on the mechanism of assigning priorities. Obviously, we can assign these priorities proportional to the wire lengths to be compensated. That is, if more wire length should be added to a wire in the next iteration, it should be assigned a large priority. This mechanism, however, may cause a slow convergence of the iterations, because wires with large priorities may not spread evenly in the routing. For example, several critical wires may compete for free spaces at one corner of the routing, while free spaces are available at another corner. This situation may happen because the inner loop L21–L32 in Algorithm 1 reduces the overall wire length to be compensated in (33), but does not consider the balance between the compensations of different wires. This situation is aggravated by the fact that there is no communication about the uncompensated wire lengths between iterations, so that the later iterations grow new meander segments from the free spaces left by the previous iterations blindly. Consequently, a more general priority assignment mechanism is required for efficient shift operations.

The basic concept of priority assignment used in our method can be explained using the example in Fig. 9c, where wires 1, 2, and 3 require no length compensation, but wire 4 needs to be compensated. In this example, the shift operation should make the free space between wires 1 and 2 available to wire 4. If we simply assign wires 1 and 2 the same priority due to the same wire lengths to be compensated, we cannot move the vertical segment on wire 2 leftwards to swap the free space to the right side. This problem can be solved by propagating the requirement of length compensation of wire 4 by increasing the priorities of the other wires 1, 2 and 3 in this example. Therefore, the priority of a wire is not only determined by its own required wire length compensation but also by the status of other wires. As the relative distance between two wires increases, the effect of priority propagation becomes weaker, so that the free space will be swapped to the direction in which more critical wires exist.

The algorithm to set priorities is summarized by the function update_priorities ($\mathcal{R}$) in Algorithm 2. In this function, $l_i$ is the wire length to be compensated in the next iteration for wire $w_i$. A larger length means that the wire has a higher priority in the next iteration. In addition, such a critical wire also increases the priorities of its neighboring wires by a reciprocal function. For example, the $j$th wire is $|i - j|$ wires away from the $i$th wire, so that the priority increase effected by $w_i$ to $w_j$ is $l_i/(1 + |i - j|)/\epsilon$. With this reciprocal relation, the length compensation status of a wire can be propagated to other wires while this wire itself still has the largest priority increase when $j = i$. This accumulative priority thus reflects the general distribution of critical wires in the routing. For example, if wires with large compensation lengths concentrated in one area, all their priorities become high due to this accumulative update.

This mechanism of setting priorities, however, can not be applied directly right after the original dense meander segments are deleted from the input routing, because wire patterns have not been tested in existing free spaces, so that the wire lengths to be compensated at this time do not reflect the real requirements of free spaces. Therefore, we only set the priorities of routing boundaries, such as board or component edges, to low priorities. These priorities are updated to the priorities of the other wires accumulatively, so that wire segments are attracted to these boundaries and hence release the center area which is competed by many wires. This initial setting is implemented in the function initialize_priorities ($\mathcal{R}$) in Algorithm 1.

**Algorithm 2:** Functions of updating priorities and shifting

- L1  $\mathcal{R}$ : a given routing;
- L2  $p_i$ : the priority of wire $w_i$;
- L3  $l_i$ : the wire length to be compensated for wire $w_i$;
- L4  $|i-j|$: index distance between wires $w_i$ and $w_j$;
- L5  $\epsilon$ : a constant scale to prevent floating point overflow.
- L6  **Function** *update_priorities* ($\mathcal{R}$)
- L7     **foreach** *wire $w_i$ in $\mathcal{R}$* **do**
- L8        **foreach** *wire $w_j$ in $\mathcal{R}$* **do**
- L9           $p_j \leftarrow p_j + l_i/(1+|i-j|)/\epsilon$;
- L10       **end**
- L11    **end**
- L12 **end**
- L13 $s_k^i$ : a wire segment on the wire $w_i$;
- L14 $s_{k,l}^i, s_{k,r}^i$ : wire segments at the left and right sides of $s_k^i$; $p_{k,l}^i, p_{k,r}^i$ : priorities of wires containing $s_{k,l}^i$ and $s_{k,r}^i$.
- L15 **Function** *shift_segments* ($\mathcal{R}$)
- L16    **repeat**
- L17       changed$\leftarrow 0$;
- L18       **foreach** *wire $w_i$ in the routing* **do**
- L19          **foreach** *straight wire segment $s_k^i$ on $w_i$* **do**
- L20             **if** $p_{k,l}^i < p_{k,r}^i$ **then** $s_m \leftarrow s_{k,l}^i$;
- L21             **else** $s_m \leftarrow s_{k,r}^i$;
- L22             **if** *moving $s_k^i$ leads to no wire length change* **then**
- L23                 move $s_k^i$ toward $s_m$;
- L24                 changed$\leftarrow 1$;
- L25             **end**
- L26          **end**
- L27       **end**
- L28    **until** *changed*=0;
- L29 **end**

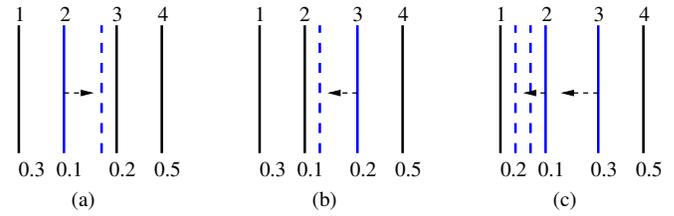

Fig. 10. Examples of shift operation based on compensation priorities. (a) Wire 2 is selected and moved toward wire 3. (b) Wire 3 is moved toward wire 2. (c) Both wire 2 and wire 3 are moved toward wire 1.

With the updated priorities, the shift operation then moves wire segments in the routing to swap free spaces. An example of this operation is shown in Fig. 10. In this operation, we select a wire segment and move it to the neighboring wire with the smaller priority. For example, in Fig. 10a we have four wire segments in solid lines with their priorities. If we check the shift condition of wire 2, we move it toward wire 3. But if we check wire 3 first as in Fig. 10b, we move it toward wire 2. However, in Fig. 10c the different orders of selecting wire 2 and wire 3 lead to the same result since we execute the shift operations repeatedly until no wire segment can be moved. Note here we shift a segment according to the priorities of its neighbors, instead of simply pushing the wire segments with large priorities to those with small priorities. If the latter mechanism were adopted, the wires would be lumped very fast and in the end we have only a bundle of wires without any floorplan information.

The iterative shift operation is summarized in the function shift_segments ($\mathcal{R}$) in Algorithm 2. In this function, we check the priorities of the neighbors of each wire segment $s_i^k$ and move it toward the segment $s_m$ that has a smaller priority. These shift operations shift_segments ($\mathcal{R}$) are executed repeatedly in each iteration in Algorithm 1 to form a new floorplan, right after priorities are initialized using initialize_priorities ($\mathcal{R}$), or refreshed using update_priorities ($\mathcal{R}$), until no change is made in the routing.

The priorities calculated by the function update_priorities ($\mathcal{R}$) contain two types of information. First, a high priority reflects that the current wire generally needs more length compensation. Second, the wires around the current wire may be critical too, because a critical wire also increases the priorities of all its neighbors. Besides using the priorities as indicators for the shift operations, we also use them to inform the solver which wires should be compensated first. Consequently, we can rewrite the objective (33) of the ILP solver as

$$\text{maximize:} \sum_{j=1}^{M_t} p_j L_j \qquad (35)$$

where we change the weights of the lengths from 1 to $p_j$. With this change, the solver tries to increase lengths of wires with large priorities, so that free spaces around critical wires will be assigned to them first. Consequently, these wires can use the available resources around them first and force other wires with lower priorities to grow new patterns in free spaces further away. This not only leads to a more balanced result from the ILP solver but also works in the same spirit of the shift operation, that is, free spaces should be made available to critical wires with high priorities.

### B. Fixing wire patterns progressively to reduce problem space

In Algorithm 1, the result of shift operations is processed in the next iteration to grow new patterns and to balance free spaces again. Because the ILP model in L22 of Algorithm 1 is created from a floorplan, we need to delete the newly generated meander segments grown in the last iteration using the function fix_remove_patterns ($\mathcal{R}_i$, $w_t$) in L34, while still keeping their compensation information by the updated priorities. This deletion can also benefit the shift operation described above. For example, in Fig. 9d, the horizontal wire segment in the middle on wire 3 might be shifted upwards and thus a very tight pattern is formed as shown with a dashed line. This pattern is also removed by fix_remove_patterns ($\mathcal{R}_i$, $w_t$), so that the free space between wires 2 and 3 becomes even larger and available to the critical wires on the right side via wire groups.

Though deleting newly grown patterns after optimization may lead to a balanced routing, a complete deletion of wire patterns is not favorable, because many patterns with good area efficiency, for example, group patterns exemplified in Fig. 3b, are also deleted and regenerated by the solver, thus increasing the convergence time of the iterative shift and optimization process. To overcome the disadvantage of complete pattern deletion, we keep some efficient patterns in the floorplan in each iteration during the execution of the function fix_remove_patterns ($\mathcal{R}_i$, $w_t$). On one hand, this will reduce the wire lengths to be compensated because we do not delete all patterns. On the other hand, the problem space of the ILP model is gradually reduced, since more and more free spaces are occupied by fixed patterns.

The fixing process starts from the edge of the routing area toward the center. The central area is usually densely populated by many wires so that a heuristic fixing may easily lead to heavy blocking. Therefore, the routing in this area is determined by the solver as a whole when the iterations finish. As the starting point, the first wire from the edge of the routing area is marked as a fixed wire after its entire length has been compensated. Thereafter, the patterns in an intermediate routing are recognized by comparing the coordinates of the bending points along adjacent wires. If the difference between the x coordinates of two bending points and the difference between the y coordinates of them are equal to $d_m$, which is the minimum distance between different wires required by manufacturing as illustrated in Fig. 6, we know that the corresponding wires are tightly routed around these bending points. We repeat this process along adjacent wires to check further bending points so that dense routing patterns similar to Fig. 3b can be identified.

After finding a wire pattern, we check whether it contains a fixed wire. For example, if wire 1 or wire 3 in Fig. 3b is a fixed wire, this wire group pattern is fixed in the following iterations. Otherwise, we remove this pattern from the intermediate routing so that the solver has a chance to regenerate new patterns in the released area later. In the iterations, wire patterns may lead to the complete length compensation of a wire that neighbors other previously fixed wires. This wire is then marked as fixed. Consequently, the overall pattern fixing proceeds gradually toward the center of the routing area.

During the iterations, it might happen that all the patterns on a wire are fixed, but the length of this wire is still not compensated completely. In this case, we delete the patterns along this wire and double its priority. Accordingly, the priorities of other wires are updated using the function update_priorities ($\mathcal{R}$) in Algorithm 2, so that the shift operations described earlier can swap free spaces to this wire.

The fixing step ensures that there is no much free space surrounded by the fixed wires. However, some patterns on the current wire may have inconvenient shapes for growing new wire patterns. For example, in Fig. 11a, there is a free space between the newly fixed wires and the wire on the right of it. Because the shift operation can only move wire segments without changing wire length, the result of shift in Fig. 11b may not occupy all the free space, in case the solver does not grow new patterns in this area. To overcome this limitation, we check the free space after running the ILP solver, and force wire 2 to bend into this area, as in Fig. 11c. The second limitation of the fixing step is that blocking patterns could be kept in the routing. For example, in Fig. 11d, the two narrow patterns block the entrance of the free space, so that wire 2 does not have access to this area. This situation might happen in the result of optimization because the solver maximizes the total compensated wire length only. To make these blocked areas available to other wires, we remove the two blocking patterns in Fig. 11d and mark wire 1 as unfixed, so that the free space is made accessible to all unfixed wires to grow compensation patterns.

The prioritized shifting and progressive fixing operations are executed in each iteration in Algorithm 1 (L20, L33, and L34). In this process, the shift operations try to move free spaces to the critical wires, and the fixing operations preserve those patterns with good area efficiency. After a few iterations, the priorities of the wires can then reflect the resource requirements, and a more balanced floorplan is thus formed. The fixation of patterns also narrow the routing area to the center area gradually, so that the routing of those wires passing this area can be evaluated quickly by the solver to achieve a balanced routing with enlarged width of dense meander segments.

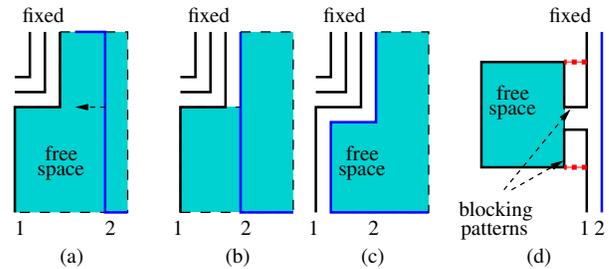

Fig. 11. Examples of fixing and blocking pattern removal. (a) The routing before shifting. (b) Simple shifting without changing wire length. (c) Forced shifting with length change. (d) Removal of blocking patterns.

## V. Experimental Results

In this section, we show experimental results of applying the proposed method on several PCB routings. The proposed framework was implemented using C++. The experiments were performed using a computer with a 2.67 GHz CPU and 8 GB memory. The solver used to solve the ILP model is Gurobi [25]. We used three test cases, case1-case3 from [14], where case3 has six buses. We have also tested the case from the BSG router in [13]. These previous methods can route given designs efficiently while matching wire lengths. However, many dense meander segments still remain in the results and are distributed unevenly. In the experiments, we set area constraints of these test cases as explained in the problem formulation in Section II, so that meander segments are generally confined by the outermost wires in the buses.

The results of the proposed method are shown in Table I, where $w_t$ is the increased width of the newly generated meander segments as defined in Fig. 1a. For example, the new width $w_t = 2$ for case1 means that the final width is two times the original one. For comparison, we also show the results from [1] in Table I. Comparing the final $w_t$ of these ten cases, we can see that nine cases out of ten have further improvement. This improvement is achieved by the introduced prioritized shifting and progressive fixing techniques explained in Section IV.

Noticeably, some test cases in Table I, for example, case3.5 and case3.6, have significant meander segment width enlargement in the comparison. In [1], meander segments can only be grown on wires requiring length compensation. Therefore, a wire without this requirement always keeps its original shape, and can thus partition the routing into two independent parts, although the free spaces in these two areas are not balanced. Consequently, the maximum achievable $w_t$ is determined by the area with fewer routing resources, while the free spaces in the other part are wasted. In the proposed shifting technique, free spaces are moved away from the wires that require little or even zero compensation, because the priority is affected not only by its own compensation but also by the compensation requirements of the other wires. Therefore, meander segments are distributed more evenly in the final routing, thus leading to a much larger width $w_t$.

TABLE I
RESULT COMPARISON WITH [1]

|  | Method in [1] |  | Proposed method |  |  |
|---|---|---|---|---|---|
|  | $w_t$ | $T(s)$ | $w_t$ | $T(s)$ | $T_{wt}(s)$ |
| case1 | 2 | 106.94 | 2 | 32.11 | 3.40 |
| case2 | 2.25 | 17.66 | 2.4 | 27.94 | 10.82 |
| case3.1 | 2.5 | 153.55 | 3 | 205.50 | 8.78 |
| case3.2 | 2 | 13.049 | 3 | 35.40 | 1.77 |
| case3.3 | 2 | 55.692 | 2.8 | 231.71 | 178.43 |
| case3.4 | 2.25 | 66.52 | 2.75 | 73.38 | 28.03 |
| case3.5 | 1.5 | 74.23 | 3 | 86.26 | 14.60 |
| case3.6 | 1 | 25.40 | 2.6 | 67.23 | 18.04 |
| case3.6s | 1 | 25.40 | 1.4 | 107.45 | 33.85 |
| case4 | 1.5 | 1510.58 | 1.6 | 3019.19 | 594.68 |

TABLE II
RESULTS OF PROGRESSIVE ROUTING REFINEMENT

|  | $1^{st}$ iteration |  |  | $2^{nd}$ iteration |  |  | Final iteration |  |  |  |
|---|---|---|---|---|---|---|---|---|---|---|
|  | $\eta_c$ | $\tau_{b,1}$ | $\tau_{c,1}$ | $\eta_{b,2}$ | $\tau_{b,2}$ | $\tau_{c,2}$ | $\eta_{b,n}$ | $\lambda$ | $\eta_x$ | $\eta_a$ |
| case1 | 17 | 1898 | 0 | 0 | 0 | 0 | 17 | 0 | 0 | 0 |
| case2 | 14 | 570 | 40 | 9 | 413.2 | 26.8 | 6 | 2 | 5 | 4 |
| case3.1 | 16 | 814 | 12 | 16 | 1347 | 0 | 16 | 1 | 0 | 0 |
| case3.2 | 12 | 116 | 0 | 0 | 0 | 0 | 12 | 0 | 0 | 0 |
| case3.3 | 20 | 1310 | 4 | 18 | 1154 | 33.2 | 5 | 7 | 4 | 2.14 |
| case3.4 | 16 | 804 | 75 | 16 | 808 | 0 | 16 | 1 | 0 | 0 |
| case3.5 | 16 | 650 | 2 | 16 | 946 | 17 | 14 | 3 | 2 | 0.67 |
| case3.6 | 12 | 314 | 0 | 0 | 0 | 0 | 12 | 0 | 0 | 0 |
| case3.6s | 12 | 302 | 10.4 | 3 | 82.4 | 0 | 3 | 1 | 9 | 9 |
| case4 | 36 | 10078 | 656 | 27 | 7806.8 | 203.2 | 8 | 5 | 11 | 5.20 |

Take case3.6 illustrated in Fig. 12a as an example of this new relaxation, where the two boundary wires almost need no compensation. Even though we consider the space outside of the original routing area to be available, the method in [1] cannot expand these existing meander segments due to the two boundary wires. Now with the help of the shifting technique, wires can be moved out of the tight confinement of the boundary wires for a better space assignment of inner wires. In Fig. 12b, the area constraints are rather loose, and the boundary wires are pushed outwards extremely to occupy the free spaces, so that the result bears nearly no similarity to the original routing. The results with these loose area constraints are shown as case3.6 in Table I. For comparison, we tightened the routing constraints and allowed smaller free spaces neighboring the boundary wires in the original routing. The results with these new constraints are denoted as case3.6s in Table I, where the width of meander segments is still increased to 1.4. This result is also illustrated in Fig. 12c, which keeps the general floorplan of the original routing but with a larger width of meander segments.

From Table I we can also observe that the runtimes of the proposed method are generally larger than the runtimes of the method [1], as shown by the two columns marked as $T$ in seconds. This is understandable since more iterations are required in the binary search to find the maximum feasible $w_t$. However, the runtimes do not increase drastically although we have introduced multiple iterations to shift the positions of long straight wire segments, because the progressive fixing technique reduces the problem space gradually so that the solver runs faster in later iterations. For comparison, if we stop the iterations in the proposed method when $w_t = 1.5$ is feasible, the total runtime is 1467.06 seconds, which is even slightly smaller than the runtime of the method in [1] to achieve the same width of the meander segments.

In the last column $T_{wt}$, the runtimes of determining the routing for the final $w_t$ of the test cases are shown. With such a runtime, the inner loop L19–L35 in Algorithm 1 for a given $w_t$ can be finished. For example, at the last step of the binary search, the width $w_t$ of case4 was 1.6, and the routing was finished within 594.68 seconds. This runtime is about one fifth of the overall runtime 3019.19 seconds, meaning that most of the runtime was spent to test other feasible or unfeasible widths during the binary search.

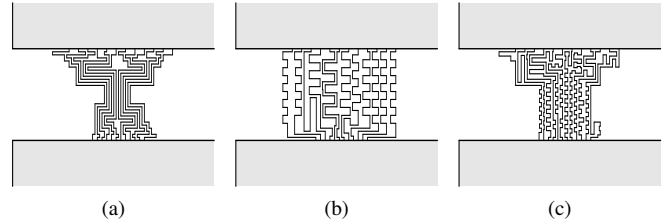

Fig. 12. Routing comparison of case3.6 and Case3.6s. (a) Original routing with $w_t = 1$. (b) Case3.6. Refined routing with loose area constraints. $w_t = 2.6$. (c) Case3.6s. Refined routing with strict area constraints. $w_t = 1.4$.

In Table II, we show more information on the prioritized shifting and progressive fixing techniques introduced in Section IV to create the new meander segments with width $w_t$ in Table I. This is the maximum feasible width found by the enhanced method in Algorithm 1. In each of the shifting and fixing iterations in Algorithm 1, the procedure from L20 to L34 is executed to move straight wire segments, update wire priorities, and fix wire patterns gradually. The meanings of columns in Table II are described as follows.

$\eta_c$: the number of wires in the bus.
$\tau_{b,1}$, $\tau_{b,2}$: the total wire length needed to be compensated before the first and the second iterations, respectively.
$\tau_{c,1}$, $\tau_{c,2}$: the total wire length still needed to be compensated after the first and the second iterations, respectively.
$\eta_{b,2}$, $\eta_{b,n}$: the number of unfixed wires before the second and the last iterations, respectively.
$\lambda$: the number of fixing iterations.
$\eta_x$: the maximum number of wires fixed in a single iteration.
$\eta_a$: the average number of wires fixed in a single iteration.

The columns of the first iteration in Table II show that three test cases case1, case3.2 and case3.6 are fully compensated with only one iteration ($\tau_{c,1} = 0$). This is because the initial priorities set in L18 of Algorithm 1 allow the shifting function shift_segments ($\mathcal{R}$) to push wire segments toward the boundaries of the routing from the beginning. For the other test cases, only small length compensations ($\tau_{c,1}$) are needed after the first iteration, when compared to the original required length ($\tau_{b,1}$). These results are considered failed and abandoned by the method in [1], where no free space can be swapped toward critical wires. In the proposed method, the compensation results are used to update wire priorities so that the given width $w_t$ may be achieved in the following iterations.

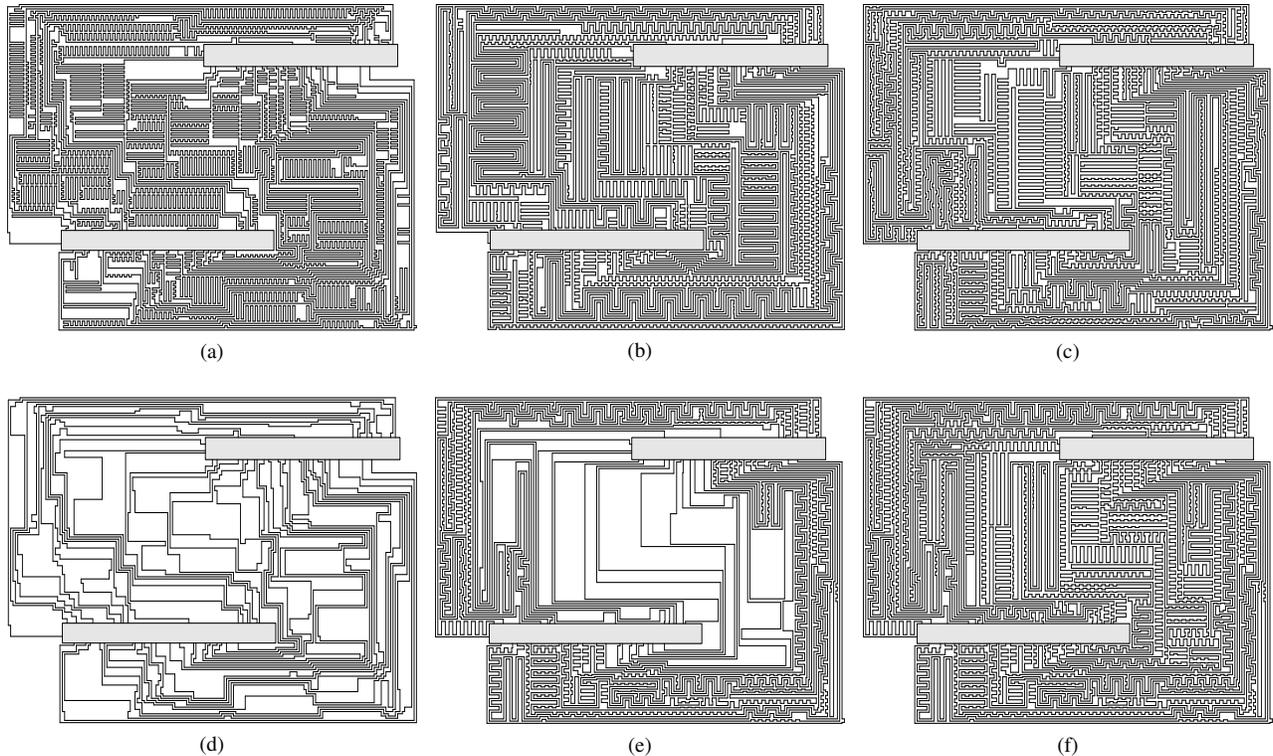

Fig. 13. Routing comparison of case4. (a) Original routing of case4 from [13]. (b) Refined routing using [1]. $w_t = 1.5$. (c) Refined routing using the proposed method. $w_t = 1.5$. (d) Floorplan right after deleting dense meander segments from the input routing. (e) Optimization input using the proposed method after the third fixation. $w_t = 1.6$. (f) Refined routing using the proposed method. $w_t = 1.6$.

In the columns of the second iteration, three more test cases case3.1, case3.4 and case3.6s finish their compensation successfully. Here $\eta_{b,2}$ is the number of wires that have not been fixed after the first iteration. $\tau_{b,2}$ represents the wire lengths to be compensated in this iteration. These values are larger than $\tau_{c,1}$ which represents the uncompensated length after the first iteration, because the meander segments that have not been fixed are removed after wire priorities are updated using the result of ILP model. Especially for case3.1, case3.4, and case3.5, the wire lengths to be compensated before the second iteration ($\tau_{b,2}$) are even larger than such lengths before the first iteration ($\tau_{b,1}$). This is because both the shifting technique and the removal of blocking patterns may also shorten wire lengths, as illustrated in Fig. 9d and Fig. 11d, respectively. For other unfinished cases, the column $\eta_{b,2}$ also demonstrates the degree of pattern fixing. Take case3.6s as example, for which 9 out of 12 wires are fixed and the total needed compensation length is decreased from 302 to 82.4, meaning a remarkable problem space reduction. Therefore, even if case3.6s cannot finish its compensation in the first iteration, later optimization iterations can be performed faster to find a feasible solution. In addition, for case3.1, case3.4 and case3.5 no wire has been fixed after the first iteration. However, the newly recorded priorities can still play an important role to guide a better free space assignment by the shifting technique, so that a balanced compensation on all wires can be achieved more effectively.

In the columns of the final iteration, $\eta_{b,n}$ shows the number of wires that have not been fixed yet before the final iteration, from which we can observe that the problem spaces in later iterations are also reduced effectively. For example, case3.3 and case4 have only 5 and 8 wires left unfixed after 7 and 5 iterations, respectively. The numbers of unfixed wires are only 25% and 22.2% of the original numbers of wires $\eta_c$, so that the corresponding ILP problems can be solved much more efficiently. Moreover, the fact that the largest $\lambda$ is only 7 in all test cases also reveals that either the prioritized shifting and optimization process is effective so that a solution with all wire lengths compensated can be obtained even when $\eta_{b,n}$ is still large, or the progressive fixing is effective so that the maximum number of wires fixed in one iteration, $\eta_x$, and the average number of wires fixed in one iteration, $\eta_a$, are large.

As an example to show the changes brought by the techniques of prioritized shifting and progressive fixing, we illustrate the refined routing of case4 with the width of meander segments as 1.5 determined by [1] in Fig. 13b, and the newly refined routing of case4 with the same width but generated by the proposed method in Fig. 13c. The original routing of case4 from [13] is shown in Fig. 13a for comparison. In Fig. 13b, the remaining free spaces spread randomly. But in Fig. 13c, the free spaces now are centralized, due to the shifting and fixing techniques, which grow efficient wire patterns starting from the edge of the routing and swap free spaces toward the center area to relieve the competition of this area by multiple wires.

In Fig. 13d-f three snapshots of the routing with meander segment width as 1.6 for case4 are shown. Fig. 13d is the floorplan right after the original dense meander segments are removed from the input routing. An intermediate result is shown in Fig. 13e where 19 out of 36 wires are fixed and the unfixed wire patterns are all erased. From this example, we can see that the fixed wire patterns generally have good routing density and only some very small free spaces still exist between wires. This figure also shows that there are many wire groups in the fixed wire patterns, because the solver tries to form wire groups as much as possible due to area efficiency. In Fig. 13f, the final routing result of case4 with the width of meander segments as 1.6 is shown. The free spaces at the center area are now much smaller compared with the result with the width of meander segments as 1.5 in Fig. 13c. The remained small free spaces in the center area hint the possibility that the width of meander segments might be enlarged further, but the improvement may be small since the width 1.75 has already been tried and is not feasible using the proposed method.

## VI. Conclusion

In this paper we have addressed the delay speedup problem caused by dense meander segments in high-performance PCBs. To extend the widths of these dense segments we proposed a post-processing framework modeling patterns in free spaces and area sharing using 0-1 variables. The problem was transformed into an ILP problem and processed by a solver to provide global balance between uncompensated wire lengths and available free spaces. Furthermore, prioritized shifting and progressive fixing are used to achieve a well-balanced routing in a reasonable runtime. Experimental results confirm that the proposed method can extend the minimum width of meander segments effectively, even to more than two times in most cases, thus reducing the speedup effect significantly.

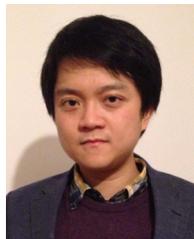

**Tsun-Ming Tseng** received the B.Sc. degree in electronics engineering from National Chiao Tung University (NCTU), Hsinchu, Taiwan, in 2010, and the M.Sc degree in communications engineering from Technische Universität München (TUM), Munich, Germany, in 2013. He is currently pursuing a Dr.-Ing. degree at the Institute for Electronic Design Automation, TUM. His current research interests include mathematical methods for computer-aided design of digital electronic circuits and microfluidic biochips.

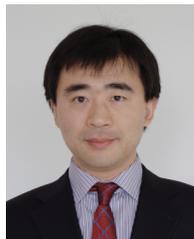

**Bing Li** received the bachelor's and master's degrees in communication and information engineering from Beijing University of Posts and Telecommunications, Beijing, China, in 2000 and 2003, respectively, and the Dr.-Ing. degree in electrical engineering from Technische Universität München (TUM), Munich, Germany, in 2010. He is currently a researcher with the Institute for Electronic Design Automation, TUM. His research interests include timing and power analysis and emerging systems.

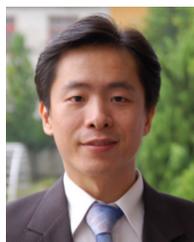

**Tsung-Yi Ho** received his Ph.D. in Electrical Engineering from National Taiwan University in 2005. He is currently a Professor with the Department of Computer Science of National Chiao Tung University, Hsinchu, Taiwan. He has been the recipient of the Invitational Fellowship of the Japan Society for the Promotion of Science (JSPS), the Humboldt Research Fellowship by the Alexander von Humboldt Foundation, and the Hans Fischer Fellow by the Institute of Advanced Study of the Technische Universität München. He currently serves as an ACM Distinguished Speaker, a Distinguished Visitor of the IEEE Computer Society, the Chair of the IEEE Computer Society Tainan Chapter, the Chair of the ACM SIGDA Taiwan Chapter, and Associate Editor of the ACM Journal on Emerging Technologies in Computing Systems and IEEE Transactions on Computer-Aided Design of Integrated Circuits and Systems, IEEE Transactions on Very Large Scale Integration Systems, and Guest Editor of IEEE Design & Test of Computers.

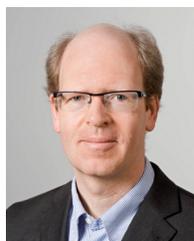

**Ulf Schlichtmann** (S'88–M'90) received the Dipl.-Ing. and Dr.-Ing. degrees in electrical engineering and information technology from Technische Universität München (TUM), Munich, Germany, in 1990 and 1995, respectively. He was with Siemens AG and Infineon Technologies AG, Munich, from 1994 to 2003, where he held various technical and management positions in design automation, design libraries, IP reuse, and product development. He has been with TUM as a Professor and the Head of the Institute for Electronic Design Automation, since 2003. He served as the Dean of the Department of Electrical Engineering and Information Technology, TUM, from 2008 to 2011. His current research interests include computer-aided design of electronic circuits and systems, with an emphasis on designing reliable and robust systems.